\documentstyle[12pt,psfig]{article} 
\newcommand{\m}{\medbreak}
\newcommand{\no}{\noindent}
\newcommand{\EQ}{\begin{equation}}
\newcommand{\eq}{\end{equation}}
\newcommand{\EQA}{\begin{eqnarray}}
\newcommand{\eqa}{\end{eqnarray}}

\newcommand{\hu}{\hspace{1.cm}}

\def\rpsq{$R_p \hspace{-1em}/\;\:$-squarks }
\def\subfigureA#1{
\leavevmode
\hbox{#1}
}
\newcommand{\appendixA}{\setcounter{equation}{0}
\def\theequation{\rm{A}.\arabic{equation}}\section*}

\def\pr#1#2#3{ Phys. Rev. {\bf{#1}}, #2 (#3)}
\def\prl#1#2#3{ Phys. Rev. Lett. {\bf{#1}}, #2 (#3)}
\def\pl#1#2#3{ Phys. Lett. {\bf{#1}}, #2 (#3)}
\def\prep#1#2#3{ Phys. Rep. {\bf{#1}}, #2 (#3)}
\def\np#1#2#3{ Nucl. Phys. {\bf{#1}}, #2 (#3)}
\def\zp#1#2#3{ Z. Phys. {\bf{#1}}, #2 (#3)}

\def\epj#1#2#3{ Eur. Phys. J. {\bf{#1}}, #2 (#3)}

\begin{document}
\begin{titlepage} 
\vspace{0.2in} 
\begin{flushright} 
\end{flushright}
\begin{center} 
{\large \bf Search and identification of Scalar and Vector
Leptoquarks at HERA with polarization \\}  
\vspace*{1.2cm} {\bf P.
Taxil{$^1$}, E. Tu\u{g}cu{$^{1,2}$} and J.-M. Virey}{$^{1,3,\star}$}\\  
\vspace*{1cm}
{$^1$}Centre de Physique Th\'eorique$^{\dag}$, CNRS-Luminy, Case 907, F-13288
Marseille Cedex 9, France and  Universit\'e de Provence, Marseille, France 
\\
{$^2$} Galatasaray University, \c  C\i ra\u gan Cad. 102, Ortak\"oy 80840-\.Istanbul, 
Turkey \\ 
{$^3$}Institut f\"ur Physik, Universit\"at Dortmund, D-44221
Dortmund, Germany\\ 
\vspace*{1.8cm} {\bf Abstract \\} 
\end{center}

We analyze the effects of Scalar and Vector Leptoquarks on
various observables in electron (positron) - proton
deep inelastic scattering. 
In view of the future program of the HERA collider, with a high luminosity and
also with polarization, we present the constraints that can be reached using this
facility for several Leptoquark scenarios. 
We address the question of the identification of the nature of
a discovered Leptoquark.
We emphasize the relevance of having polarized lepton and proton beams 
in order to disentangle completely the various Leptoquark models.
This study is also relevant in the context of the TESLA$\times$HERA project.

\vspace{0.8cm}

\vfill \begin{flushleft} PACS Numbers : 12.60.-i; 13.88.+e; 13.85.Qk; 13.85.Rm

\m\no December 1999\\ DO-TH 99/15\\ 
CPT-99/P.3831\\ \m\no

------------------------------------
\\ $^{\dag}$Unit\'e Propre de Recherche 7061

{$^\star$} Fellow of the ``Alexander von Humboldt'' Foundation, 
present address: $1$, email:
virey@cpt.univ-mrs.fr \end{flushleft} 
\end{titlepage}


\section{Introduction}

\vspace{1mm} 
\noindent 
Many extensions of the Standard Model (SM), like for instance
Supersymmetry (SUSY) or Grand Unified Theories (GUT), 
predict the existence of Leptoquarks (LQs),
which are particles that couple directly to quark-lepton pairs. 
In general there is no particular prediction for the masses of these LQs, 
which can range
from the electroweak (EW) scale to the GUT scale. 
However an interesting possibility
is the case of SUSY models where the $R_{Parity}$ symmetry \cite{Fayetrp}
is violated (for a recent review on this subject see \cite{GDR}).
Then some \rpsq have direct couplings to electron-quark pairs and
are completely analogous to some of the LQs considered here.
The equivalence between \rpsq and LQs is described, for instance, in \cite{kalino}.
An interesting feature of the SUSY models with \rpsq is that the squarks
could have some relatively low masses (between the EW and the TeV scales)
since SUSY is believed to be broken at the TeV scale.\\

In this paper we will not consider any precise \rpsq model but we rather
adopt the ``model independent'' approach  of Buchm\"uller-R\"uckl-Wyler 
(BRW)\cite{BRW}, 
where the LQs are classified according to their quantum numbers 
and have to
fulfill several assumptions like $B$ and $L$ conservation (to avoid rapid proton
decay) and $SU(3)$x$SU(2)$x$U(1)$ invariance. We refer to \cite{BRW}
for more details.
The interaction lagrangian for scalar leptoquarks is given by : 
\EQA
{\cal{L}}_{scal}&=&\left(g_{1L}\, \bar{q}_{L}^ci\tau_2\ell_L+g_{1R}\, 
\bar{u}_{R}^ce_R \right)\, .\,
{\bf S_1}\, +\, \tilde{g}_{1R}\, \bar{d}_{R}^ce_R \, .\, {\bf \tilde{S}_1}\, +\,
g_{3L}\, \bar{q}_{L}^ci \tau_2{\bf \tau}\ell_L\, .\, {\bf S_3} \nonumber\\ &+&\;
\left(h_{2L}\, \bar{u}_R\ell_L+h_{2R}\, \bar{q}_Li\tau_2 e_R\right)\, .\, 
{\bf R_2}\, +\,
\tilde{h}_{2L}\, \bar{d}_R\ell_L\, .\, {\bf  \tilde{R}_2}\;\; , \eqa

\no where the scalar LQs $S_1$, $\tilde{S}_1$ are singlets and $S_3$ is a triplet,
all with fermionic number ($F=3B+L$) $F=2$. 
$R_2$ and $\tilde{R}_2$ are  doublets with $F=0$.
$\ell_L$, $q_L$ ($e_R$, $d_R$, $u_R$) are the usual lepton and
quark doublets (singlets). \\
For vector LQs the lagrangian is :
\EQA
{\cal{L}}_{vect}&=&\left(h_{1L}\, \bar{q}_{L}\gamma^{\mu }\ell_L+h_{1R}\, 
\bar{d}_{R}\gamma^{\mu}e_R \right)\, .\,
{\bf U_{1\mu }}\, +\, \tilde{h}_{1R}\, \bar{u}_{R}\gamma^{\mu }e_R \, .\, 
{\bf \tilde{U}_{1\mu }}\, +\,
h_{3L}\, \bar{q}_{L}\tau\gamma^{\mu }\ell_L\, .\, {\bf U_{3\mu }} \nonumber\\ &+&\;
\left(g_{2L}\, \bar{d}^c_R\gamma^{\mu }\ell_L+g_{2R}\,
 \bar{q}^c_L\gamma^{\mu }e_R\right)\, 
.\, {\bf V_{2\mu }}\, +\,
\tilde{g}_{2L}\, \bar{u}^c_R\gamma^{\mu }\ell_L\, .\, {\bf  \tilde{V}_{2\mu }}\;\;
 , \eqa

\no where the vector $U_{1\mu }$, $\tilde{U}_{1\mu }$ are singlets and $U_{3\mu }$ 
is a triplet,
all with $F=0$, and
$V_{2\mu }$, $\tilde{V}_{2\mu }$ are  doublets with $F=2$.\\
\no Therefore, if one takes into account the left and right-handed chiralities
${\cal{L}}_{scal} + {\cal{L}}_{vect} \  $ yields 14 independent models of LQs.
From these two lagrangians one can deduce some properties of the LQ models which 
are compiled in Table 1 of \cite{kalino}.  A point which is important to notice is that the LQ
couplings are flavor dependent.
In what follows we denote generically by $\lambda$ any LQ
coupling and by $M$ the associated mass. \\
In addition, in order to simplify the analysis, we make the following assumptions : 
{\it i}) the LQ
couples to the first generation only, {\it ii}) one LQ multiplet is present at a time,
{\it iii}) the different LQ components within one LQ multiplet are mass degenerate,
{\it iv}) there is no mixing among LQs. \\

The LQs are severely constrained by several different experiments, and we refer to
\cite{david,leurer,LQlim} for some detailed discussions. Here we only quote the most important
facts : \\

{\small 1)} Leptonic pion decays and $(g-2)_\mu$ measurements indicate that the LQs must be
chiral \cite{david,leurer} ({\it e.g.} $S_1$ and $R_2$ could have Left-handed or Right-handed 
couplings but not both). \\

{\small 2)} 
To avoid the stringent constraints from FCNC processes the simplest
assumption is to impose ``family diagonal'' couplings for the LQs, namely they couple to 
only one generation \cite{david}.\\

{\small 3)} There are some collider constraints coming from Tevatron through
the searches for LQ pair production. This process, which involves the color
properties of the LQs, yields some bounds on the mass of the LQs 
independently of the $\lambda$ coupling and of the particular scalar or vector LQ model. 
However, these mass bounds are strongly
dependent on the branching ratio $BR(LQ \rightarrow eq)$, and the values quoted below
correspond to the  maximal case $BR=1$.\\
For the scalar LQ models, the CDF and D0 collaborations at the Tevatron have combined
their data to provide \cite{LLCWG} the constraint : $M>242\, GeV\; (BR=1)$.
The dependence of these limits on $BR$ is presented in \cite{LLCWG}.\\
For the vector LQ models the situation is more complex since, in general, the experimental
bounds depend on two new parameters : $\kappa_g$ and $\lambda_g$. These two parameters
correspond to the possible anomalous couplings present at vertices involving
gluon(s) plus vector LQ(s) \cite{blumboos}.
The value of the cross sections depends on these two parameters. In particular,
the smallest cross sections do not correspond in general to the ones
obtained for ``pure'' gauge boson couplings (i.e. $\kappa_g=0$, $\lambda_g=0$).\\
The D0 collaboration has published some mass bounds for vector LQs for several values of 
($\kappa_g$,$\lambda_g$) \cite{D0LQv}. They obtained $M>340\, GeV\; (BR=1)$ for
($\kappa_g=0$,$\lambda_g=0$), but the weakest constraint, corresponding to the minimal
cross section, $M>245\, GeV\; (BR=1)$ is obtained for ($\kappa_g=1.3$,$\lambda_g=-0.2$).\\
To conclude this part, we remark that the minimal mass bound for vector LQs is close
to the mass bound for scalar LQs, and we recall that these bounds are strongly dependent on $BR$.
For instance, the \rpsq models mentioned above are not coupled to $e-q$ pairs only
but also to some superpartners ($R_p$ conserving decays)
which means that $BR < 1$. 
As a consequence, for some particular models, $BR$ can be relatively small,
giving much lower LQs mass limits.\\

{\small 4)} Low Energy Neutral Current data, in
particular from Atomic Parity Violation on Cesium atoms (APV) 
experiments, give in general the strongest bounds on the ratio $M/\lambda$.
In fact, the last experimental results on the measurements of
$Q_W(Cs)$, the weak charge for Cesium atoms, give \cite{APVexp}:
$Q_W^{exp}=-72.06 \pm (0.28)_{exp} \pm (0.34)_{th}$. For the SM we expect 
\cite{PDG}: $Q_W^{th}=-72.84 \pm 0.13$. 
This means that the SM is excluded at the
1.8$\sigma$ level. However this discrepancy is not huge and even if the
experimental errors have strongly decreased compared with preceding experiments, 
they are
still sizable. 
Then we take these results with some caution.\\
Nevertheless, we have used the formula from \cite{leurer} 
to compute the constraints on the LQ models
taking these new data into account.

In fact, since the $Q_W$ experimental value does not exactly correspond to 
the SM prediction, we
need some new physics effects to fit the data. 
Consequently, on the one hand,
if a particular LQ model gives a deviation for $Q_W$ which is in the wrong
direction with respect with the measured value, then it is simply
excluded for any value of $M/\lambda$. On the other hand, if the deviation of the 
LQ model
is on the right direction, this LQ model helps to fit the data and we get not
only an upper  bound for $M/\lambda$ but a window of ``presence'', namely
the LQ should have a ratio $M/\lambda$ in this window to make agreement 
between data and theory. The figures we obtain for the BRW LQ models considered here 
are 
given in Table 1 : 
\begin{table}[htbp]
\begin{center}
\begin{tabular}{|c|c|c|c|}
\hline
{\bf Leptoquark}	&{\bf Limits}&{\bf Leptoquark}	&{\bf Limits} \\
\hline
$S_{1L}$	&	1600-3900   & $U_{1L}$	&	$\times$ \\
\hline
$S_{1R}$ &		$\times $  & $U_{1R}$	&	2400-5800  \\
\hline
$\tilde{S}_{1R}$ &	$\times$   & $\tilde{U}_{1R}$ &	2300-5500 \\
\hline 
$S_{3L}$	&	2900-7000   & $U_{3L}$	&	$\times$  \\
\hline
$R_{2L}$	&	$\times$   & $V_{2L}$	&	2400-5800  \\
\hline
$R_{2R}$	&	2350-5650   & $V_{2R}$	&	$\times $ \\
\hline
$\tilde{R}_{2L}$ &	$\times $  & $\tilde{V}_{2L}$  &	2300-5500  \\
\hline
\end{tabular}
\end{center}
\end{table}
\vspace*{-0.8cm}
\begin{center} Table 1: Limits on $M/\lambda$ in
$GeV$ at 95\% CL from APV. 
\end{center}

\no In this table a cross indicates that the model is excluded.\\

Note that similar bounds have been obtained for LQ models from the GUT group $E_6$  
\cite{casalLQ}. These constraints can be relaxed
if there are some compensating contributions coming from more than one source of new
physics \cite{Bargerct3}.\\ 
Adopting a conservative attitude we do not consider the last measurement 
of $Q_W$  as a clear evidence for 
new physics, and in the following we will consider that 
{\it all} the LQ models can still
exist at low energy scales and can induce some effects 
in deep inelastic scattering (DIS).\\

{\small 5)} Finally, there are also some collider constraints coming from LEP
 \cite{LEPLQ} 
and HERA \cite{Perez}. In fact, depending on the particular LQ model involved the limits
obtained at these facilities cover in general a small part of the parameter space 
($M$,$\lambda$). 
\medbreak
The analysis of LQ
effects at present or future $ep$ machines is of particular relevance since such
particles could be produced in the $s$-channel \cite{BRW}.\\
In this paper, we complete and extend
the analysis which has been presented recently \cite{VTT} on the effects of Scalar LQs
in the Neutral Current (NC) and Charged Current (CC) channels at HERA. 

We will concentrate on the HERA collider with high integrated luminosities
and also  with a slightly higher energy in the center of mass. Namely we take
$\sqrt{s} = 380\, GeV$ in order to increase the domain of sensitivity for the 
LQ models.
This value for the energy could be reached in the future at HERA \cite{botje}. 
However,
we consider also the case $\sqrt{s} = 300\, GeV$ in order to test the impact of
the energy value on the capabilities of the HERA collider to discover LQs. 
In addition,
we are also concerned with a possible new $ep$ collider running at much higher 
energies $\sqrt{s} = 1\, TeV$, like the TESLAxHERA project \cite{Sirois}.\\

An other important point of our analysis is that 
we consider the case where polarized beams are available.  
Indeed, thanks to the progress which
have been performed by the RHIC Spin Collaboration \cite{RSC} at Brookhaven, the
acceleration of polarized proton beams up to high energies is becoming a true
possibility. Adding this opportunity to the fact that high
intensity polarized lepton beams will certainly be available soon at HERA, and also at
a future linear accelerator, some new windows could be opened with
$\vec{e}\vec{p}$ collisions.
The resulting potentialities for HERA physics have been discussed in several
recent workshops \cite{hera96,hera97,hera98,hera99}.\\

This paper is organized as follows :
In section 2, we estimate the constraints on the parameter space that can be 
reached 
in the future at HERA for
several Leptoquark scenarios and we compare these results with the 
present bounds.
In section 3, we propose a strategy for the identification of 
the various LQ models.
In particular, we show that both electron and proton polarizations
($\vec{e} + \vec{p}$) are necessary
to disentangle the different models.
Finally we summarize our results and we conclude in section 4.
The details of the formulas we have used are given in the Appendix.\\

\section{Discovery limits from future $ep$ experiments}

We consider the HERA collider with ${e^-}$ or ${e^+}$ beams
but with some high integrated luminosities, namely
$L_{e^-}=L_{e^+}=500\, pb^{-1}$. 
The other parameters for the analysis being \nolinebreak :
$\sqrt{s}=380\, GeV$, $0.01<y<0.9$, 
$\left( \Delta\sigma /\sigma \right)_{syst} = 2\, \%$
and we use the GRV partonic distribution functions (pdf) set \cite{GRV}. \\

We present in Fig. 1
the discovery limits at 95\% CL for the various
 LQ models 
that we obtain from a $\chi^2$ analysis performed on
the unpolarized NC cross sections 
$d\sigma/dQ^2$ for $ep \rightarrow eX$ at leading order 
(see the Appendix). \\
Next-to-leading order (NLO) QCD corrections to the production cross-section
have been estimated recently \cite{Plehn,KS}. In the mass range we consider, 
$K-$factors increase the cross section by up to 
30-50\% according to two different calculations \cite{Plehn,KS}. This means that
our bounds are somewhat pessimistic. On the other hand, we expect
that the asymmetries we will present later will be less affected
by NLO corrections since $K-$factors should cancel in the ratios.\\
\begin{figure}[ht]
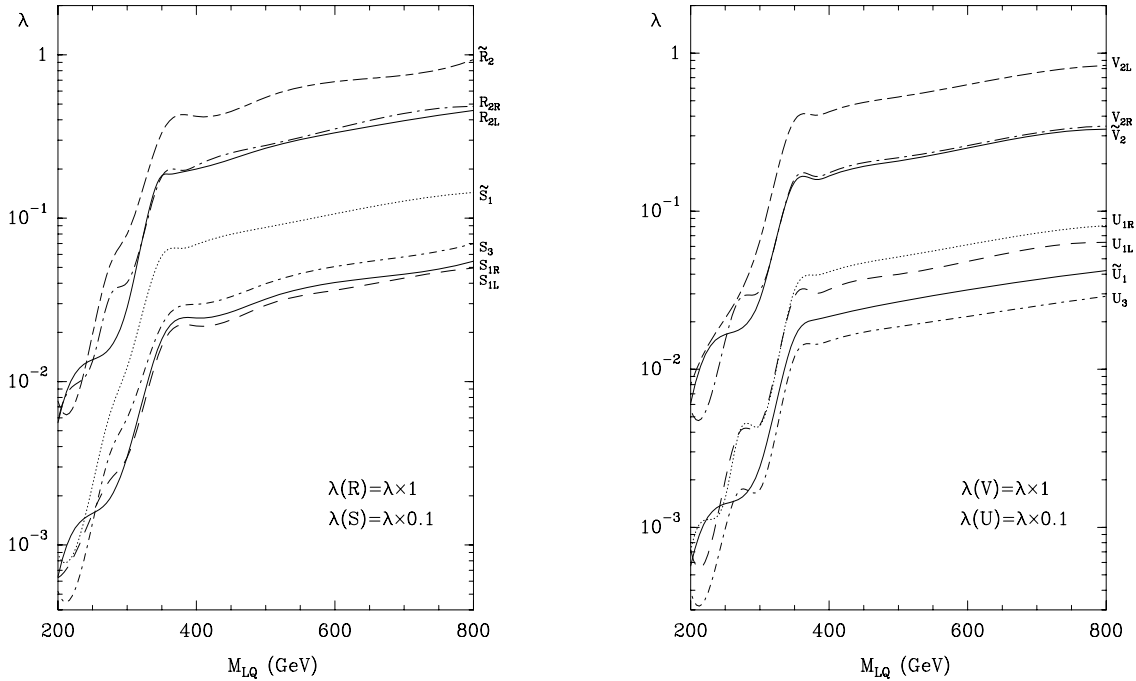

\vspace*{-2.2cm} 
\begin{tabular}[t]{c c}
\centerline{\subfigureA{\psfig{file={fig.1a},width=8truecm,height=12truecm}}
\subfigureA{\psfig{file={fig.1b},width=8truecm,height=12truecm}}}
\end{tabular} 
\vspace*{-1.6cm}
\caption{\small Discovery limits at 95\% CL for the various
LQ models at HERA.} 
\end{figure}

From these plots we see that  there are clearly two 
different domains of constraints in the plane ($M,\lambda$). 
In the "real
domain", ($M<380\, GeV$), production in the $s$-channel is by far dominant due to
 the resonance.
The "virtual domain" for masses above 380 $GeV$ corresponds to the production or 
exchange of an
off-shell LQ and the SM cross-section is less affected. 
As a consequence the bounds are weaker. 
In addition we see on this figure that the LQs which couple preferentially to 
$d$-quarks
 ($\tilde{S}_{1}$,
$\tilde{R}_{2}$) and (${V}_{2L}$,${U}_{1R}$) are less constrained compared to the 
others since $u$
quarks are dominant in the proton. 
Isospin symmetry implies that we would need $en$ collisions 
(with the same values for $\sqrt{s}$ and $L$) to
constrain these two LQs at the same level. \\

Besides the discovery bounds obtained from the
unpolarized NC cross sections, it is interesting to examine which sensitivity
could be obtained from other observables like the unpolarized CC cross sections,
the single or double polarized cross sections (in NC or CC processes)
or some spin asymmetries.\\

\no $\bullet $ {\it Charged Current cross sections :} 
\vspace{2.mm}

Concerning the CC channel, in DIS the SM process corresponds to
$W$ exchanges : $ep \rightarrow \nu X$. The LQs which have both couplings to
$eq$ and $e\nu$ pairs should also induce some effects in CC processes. 
Note that within our assumptions (no mixing) only $S_{1L}$ , 
$S_3$, $U_{1L}$ and  $U_3$ could induce some effects in
the CC sector.\\ 

The effects in  CC at HERA have been analyzed some time
ago in \cite{DonHew} in the framework of some specific models
based on superstring-inspired $E_6$. More recently they were 
considered again, essentially in the 
context of the so-called HERA anomaly problem (for a review and references, 
see \cite{Altarelli}). 
Then, the CC process was considered
to analyze the origin of the LQ rather than in the purpose of discovery.
Here we can confirm this strategy since 
our $\chi^2$ analysis shows that the sensitivity of the CC 
unpolarized cross section to the presence of the LQs is 
well below the one of the NC unpolarized cross section.
Therefore, in the following of this section we do not consider anymore 
the CC processes.\\

\no $\bullet $ {\it Polarized cross sections :} 
\vspace{2.mm}

When polarized beams are available, the basic observables are 
the helicity-dependent
(=polarized) cross sections. Polarization adds several types of
systematic errors to the unpolarized case (see \cite{syst} for example).
Then, in general, with polarized beams one prefers
to use some spin asymmetries rather than the individual polarized cross sections.
Indeed, most of the systematics cancel in the differences between 
the cross sections in various helicity states (numerator) and also in the ratio.
However, at HERA, one can expect relatively small systematics
for the polarized cross sections themselves.
In particular one expects \cite{deR2}: 
\EQ
{\large(}{{\Delta \sigma^{h_e}} 
\over \sigma^{h_e}}{\large )}_{syst} = 2 - 5 \% 
\;\;\;\;\;\;\;\;\;
and \;\;\;\;\;\;\;\;\;
{\large(}{{\Delta \sigma^{h_e,h_p}} 
\over \sigma^{h_e,h_p}}{\large )}_{syst} = 5 - 10 \% 
\eq
\no where $h_e,h_p$ are the helicities of the electron and of 
the proton (protons are not polarized in the first case).

Therefore, we have computed the sensitivities of the polarized cross sections
using the most favorable values for the systematics. For the calculations we have
assumed a degree of polarization $P=70\%$  and used the GRSV polarized pdf set
\cite{GRSV}. 

In comparison with the unpolarized NC cross sections we find, on the one hand,
that 
the double polarized NC cross sections have a sensitivity of the
same order and, on the other hand, that
the single polarized NC cross sections have a slightly better sensitivity by
roughly $2-10\; \% $ (the precise value depending on the model). These conclusions
are only indicative because the sensitivity of the cross sections (polarized or not)
are strongly dependent on the systematics.\\

\no $\bullet $ {\it Spin asymmetries :} 
\vspace{2.mm}

Finally, we have also computed the constraints that can
be reached by studying some Parity Violating (PV) spin asymmetries (definitions below). 
Concerning the systematic errors, we have considered $(\Delta A/A )_{syst} = 10\; \%$
which is the expected value \cite{syst}.
It appears that when both lepton and proton beams are polarized, the
limits are very close to the ones obtained in
the unpolarized case. When  lepton polarization only is
available the bounds are slightly weaker. \\

In conclusion, for the purpose of discovery
the most simple way to proceed at HERA is to consider the NC unpolarized 
cross sections.\\


In figure 2 we present a comparison
in the plane ($M,\lambda$) between the present constraints and
what could be achieved in the future in $ep$ collisions.

\begin{figure}
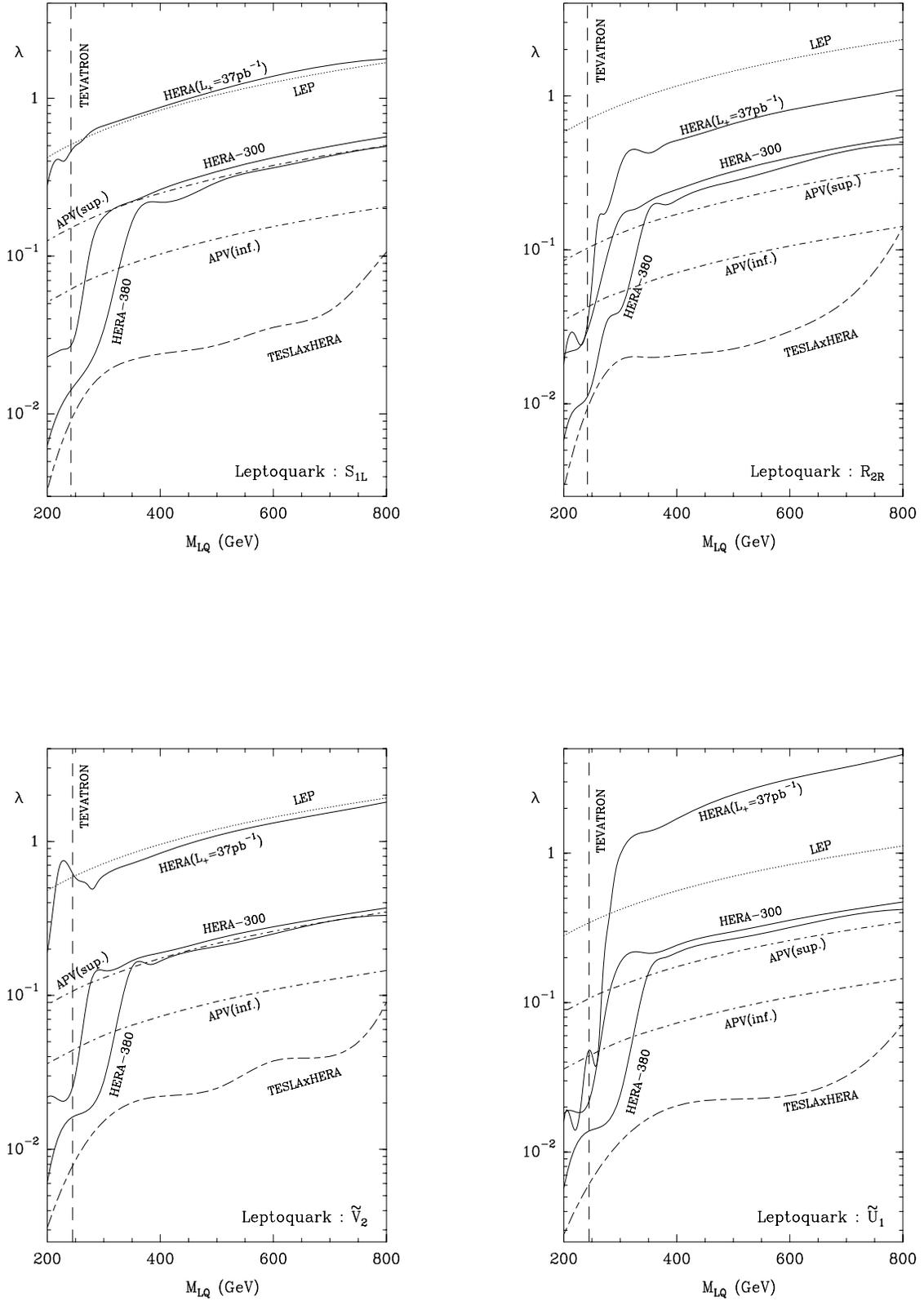

\vspace*{-2.2cm} 
\begin{tabular}[t]{c c}
\centerline{\subfigureA{\psfig{file={fig.2a},width=8truecm,height=12truecm}}
\subfigureA{\psfig{file={fig.2b},width=8truecm,height=12truecm}}}\\
\centerline{\subfigureA{\psfig{file={fig.2c},width=8truecm,height=12truecm}}
\subfigureA{\psfig{file={fig.2d},width=8truecm,height=12truecm}}}
\end{tabular} 
\vspace*{-1.4cm}
\caption{\small Constraints at 95\% CL for various
present and future experiments for ${R}_{2R}$,
${S}_{1L}$, $\tilde{V}_{2}$ and $\tilde{U}_{1}$.} 
\end{figure}

\no
We have shown the cases of two different scalars and two different vectors
for illustration. The situation is very similar for the 10
remaining models.
For future experiments one considers, 
on the one hand, the HERA collider with a higher energy of
$\sqrt{s}=380\ GeV$ but also with  $\sqrt{s}=300\ GeV$
and, on the other hand, the very interesting project
TESLA(e)xHERA(p) where an energy of $\sqrt{s}=1\, TeV$ could be reached 
\cite{Sirois}.
The integrated luminosities are 
$L_{e^-}=L_{e^+}=500\, pb^{-1}$ in all cases.\\

We can remark the followings :
{\small 1)} For most of the models, LEP limits are already covered 
by present HERA data \cite{Perez}. 
{\small 2)} Concerning the constraints from
APV  experiments, we show the allowed windows which are obtained
by taking seriously into account the recent results on $Q_W$ and their
interpretation in terms of a New Physics due to a LQ. Then,
in the virtual domain, the expected sensitivity of the future HERA
program would not give better insights than the APV experiments with
their present sensitivity. In the real domain the situation is different.
{\small 3)} Tevatron data cover an
important part of the parameter space in the real domain. 
However, we recall that the bounds obtained
from LQ  pair production at Tevatron are strongly sensitive to  $BR(LQ \rightarrow
eq)$. This is the case for R-parity
violating squarks in SUSY models \cite{kalino}. 
{\small 4)} To increase the window of
sensitivity in the real domain, it is more important to increase the energy than the
integrated luminosity. 
{\small 5)} The TESLAxHERA project will give access to a domain (both
real and  virtual) which is unconstrained presently. However, if this project is 
achieved, it will run at a time where the LHC will be running too. Then, there will be
some important constraints on $M$ from LQ pair production at LHC, 
but again those
constraints will be strongly model dependent ({\it i.e.} $BR$ dependent). \\

We conclude that there are still some windows for discovery at HERA and at
future $ep$ machines, in complementarity with the constraints coming from LQ pair
production at pure hadronic colliders.
We now turn to the problem of the identification of the nature of the LQ, a problem
which is much more difficult and where polarization will be of great help.\\

\section{Strategy for the identification of the various LQ models}

\subsection{Observables in a future HERA program}

Besides the unpolarized differential cross sections $d\sigma_{\pm}/dy$
and  $d\sigma_{\pm}/dQ^2$ in both the $e^+$ and $e^-$ channels, we have 
considered
a large set of polarized observables like the spin asymmetries.
Indeed, since the LQs are chiral one can expect that the most important 
effects will appear on the Parity
Violating (PV) spin asymmetries which can be defined when both beams 
are polarized or
when there is lepton polarization only. Parity Conserving (PC) spin asymmetries
will also be of great help as well as some charge asymmetries.

We will only define and discuss below the quantities which turned out to
be the best ones to pin down the nature of the LQ and which have the
stronger sensitivity to this kind of new physics.
We will start by recalling the definitions of the relevant asymmetries.\\

If one beam is polarized (in practice, the lepton beam) one can define
the single-spin parity-violating longitudinal asymmetry  $A_{L}(e^t)$ :
($t = \pm$ according to the electric charge of the lepton)
\EQ\label{defAL}
 A_{L}(e^{t})\; =\;
\frac{\sigma^{-}_{t} \, -\, \sigma^{+}_{t}} {\sigma^{-}_{t} \, +\,
\sigma^{+}_{t}} \;\; , 
\eq 
\no 
where $\sigma_{t}^{h_e} \equiv 
(d\sigma_{t}/dQ^2)^{h_e }$ and  $h_e$ is the
helicity of the lepton. 
In addition, when both lepton and proton beams
are polarized, some double-spin PV asymmetries can be defined  \cite{JSJMV}.
For instance $A_{LL}^{PV}$ is defined as : 
\EQ\label{defALLPV}
 A_{LL}^{PV}(e^{t})\; =\;
\frac{\sigma^{--}_{t} \, -\, \sigma^{++}_{t}} {\sigma^{--}_{t} \, +\,
\sigma^{++}_{t}} \;\; , 
\eq 
\no 
where $\sigma_{t}^{h_e h_p} \equiv 
(d\sigma_{t}/dQ^2)^{h_e  h_p}$, and $h_e, h_p$ are the
helicities of the lepton and  the proton, respectively.\\

On the other hand, with longitudinally polarized beams,
one needs two polarizations to define
some parity-conserving (PC) asymmetries $A_{LL}^{PC}$. 
These well-known quantities have been extensively
used in polarized DIS to determine the spin structure of the 
nucleon \cite{Reya}.\\
Here we will use the following :
\EQ\label{defA1}
 A_{1}^{PC}\; =\;
\frac{\sigma^{--}_{-} \, -\, \sigma^{-+}_{-}} 
{\sigma^{--}_{-} \, +\, \sigma^{-+}_{-}} 
\;\; , 
\eq 
\no 
\EQ\label{defA2}
 A_{2}^{PC}\; =\;
\frac{\sigma^{++}_{-} \, -\, \sigma^{+-}_{-}} 
{\sigma^{++}_{-} \, +\, \sigma^{+-}_{-}} 
\;\; , 
\eq 
\no  and 
\EQ\label{defA3}
 A_{3}^{PC}\; =\;
\frac{\sigma^{++}_{+} \, -\, \sigma^{+-}_{+}} 
{\sigma^{++}_{+} \, +\, \sigma^{+-}_{+}}  
\;\; , 
\eq

Finally, since  $e^-$ as well as $e^+$ (polarized) beams will be available
at HERA, one can define
a large set of (polarized) charge asymmetries \cite{JMV}. Among this set, only the
following turned out to be relevant for our purpose :
\EQ\label{defBu}
\hspace*{-0.5cm}
 B_{U}\; =\;
\frac{\sigma^{--}_{-} - \sigma^{++}_{-} + \sigma^{++}_{+} - \sigma^{--}_{+} +
\sigma^{-+}_{-} - \sigma^{+-}_{-} + \sigma^{-+}_{+} - \sigma^{+-}_{+}}
{\sigma^{--}_{-} + \sigma^{++}_{-} + \sigma^{++}_{+} + \sigma^{--}_{+} +
\sigma^{-+}_{-} + \sigma^{+-}_{-} + \sigma^{-+}_{+} + \sigma^{+-}_{+}} 
\; =\; 
\frac{\sigma^{-0}_{-} - \sigma^{+0}_{-} + \sigma^{0+}_{+} - \sigma^{0-}_{+}}
{\sigma^{-0}_{-} + \sigma^{+0}_{-} + \sigma^{0+}_{+} + \sigma^{0-}_{+}}
\;\; , 
\eq 
\no
and
\EQ\label{defBv}
\hspace*{-0.5cm}
 B_{V}\; =\;
\frac{\sigma^{--}_{-} - \sigma^{++}_{-} + \sigma^{--}_{+} - \sigma^{++}_{+} +
\sigma^{+-}_{-} - \sigma^{-+}_{-} + \sigma^{-+}_{+} - \sigma^{+-}_{+}}
 {\sigma^{--}_{-} + \sigma^{++}_{-} + \sigma^{--}_{+} + \sigma^{++}_{+} +
\sigma^{+-}_{-} + \sigma^{-+}_{-} + \sigma^{-+}_{+} + \sigma^{+-}_{+}}
\; =\; 
\frac{\sigma^{0-}_{-} - \sigma^{0+}_{-} + \sigma^{-0}_{+} - \sigma^{+0}_{+}}
{\sigma^{0-}_{-} + \sigma^{0+}_{-} + \sigma^{-0}_{+} + \sigma^{+0}_{+}}
\;\; , 
\eq 
\no where the index $0$ means unpolarized and the order $h_{e},h_{p}$
has been respected. Note that both lepton and proton polarizations 
are necessary if one wants to measure these quantities.

\subsection{Unpolarized case}
We consider first the case of neutral currents.\\
If a LQ is present in an accessible kinematic range at HERA, 
it will be discovered from the analysis of $d\sigma_{t}/dQ^2$
which have the greatest "discovery" potential. However,
if one starts trying to pin down the various models, both 
$d\sigma_{t}/dy$ and  $d\sigma_{t}/dQ^2$ have to be analyzed 
simultaneously.\\

As is well known \cite{BRW} the use of $e^-$ or $e^+$ beams allows
the separation of the 14 models of LQs into two classes according to
the value of the fermionic number F. This comes from the dominant 
(LQ mediated) interaction
between a valence quark and an $e^-$ (F = 2) or an $e^+$ (F = 0).

Hence, a deviation from $d\sigma^{SM}_{-}/dQ^2$ indicates the 
class ($S_{type}\; or\; V_{type}$), whereas a deviation from 
$d\sigma^{SM}_{+}/dQ^2$ corresponds to
the class ($R_{type}\; or\; U_{type}$).\\ 

Then, the $y$ dependence, which is obtained from 
the two $d\sigma_{t}/dy$, is the best way to discriminate between a scalar and
a vector interaction. Indeed, the SM background displays
$d\sigma_{t}/dy \, \sim 1/y^2$  when the pure vector LQ case goes as
$y$ and the pure scalar LQ is constant in $y$. It is straightforward to obtain 
these behaviours from the formulas given in \cite{BRW,kalino}
and in the Appendix.\\
We illustrate this pattern in Fig. 3 for two different choices
 of scalar and vector
LQs, with parameters allowed by the present limits. 
\begin{figure}
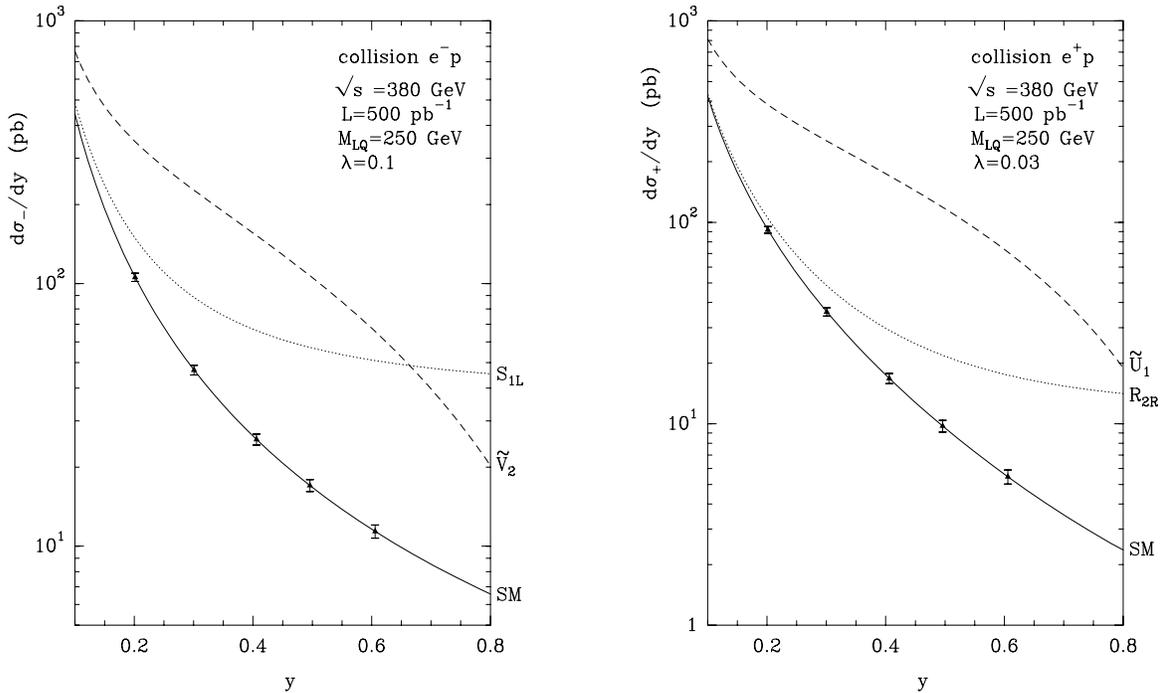

\vspace*{-2.2cm} 
\begin{tabular}[t]{c c}
\centerline{\subfigureA{\psfig{file={fig.3a},width=8truecm,height=12truecm}}
\subfigureA{\psfig{file={fig.3b},width=8truecm,height=12truecm}}}
\end{tabular} 
\vspace*{-1.4cm}
\caption{\small  $d\sigma_{-}/dy$ for 
${S}_{1L}$ and $\tilde{V}_{2}$, and $d\sigma_{+}/dy$ for ${R}_{2R}$ and $\tilde{U}_{1}$.} 
\end{figure}
Since the separation is easy, in the following we will treat scalar and
vectors as two distinct species. Now the LQ models are separated in four 
distinct classes :($S_{type}$), ($R_{type}$), ($V_{type}$) and ($U_{type}$).\\

On the other hand, Charged Current (CC) processes could in 
principle allow to go further into the distinction procedure.
We have seen previously that only $S_{1L}$
and $S_3$ for scalars, $U_{1L}$
and $U_3$ for vectors, can induce a deviation from SM expectations (if we do not 
assume LQs mixing\footnote{We refer to \cite{BKMRCC} for a discussion on scalar
 LQs mixings.}).
This means that the analysis of $\sigma^{CC}_{e^-p}$ allows to split the 
($S_{type}$) class into
($S_{1L}$,$S_3$) and ($S_{1R}$,$\tilde{S}_{1}$), while the ($U_{type}$) class is
split into ($U_{1L}$,$U_3$) and ($U_{1R}$,$\tilde{U}_{1}$).

In addition, it appears that when LQ exchange interferes 
with $W$ exchange, $S_{1L}$ and $S_3$ display some opposite patterns
(see Appendix), and this is also the case  between $U_{1L}$ and $U_3$.
However this interference term is too small to be
measurable from unpolarized CC processes at HERA 
within the allowed parameters. \\

Then, if we want to go further into the identification of the LQs we need to
separate "$eu$" from "$ed$" interactions, which seems to be impossible within $ep$
collisions except if the number of anomalous events is huge \cite{Sirois2}. 
If $en$ collisions were available, the
analysis of the respective $ep$ and $en$ production rates should 
allow this separation \cite{VTT}.\\

In conclusion, the  $ep$ unpolarized studies should allow
the separation of the 14 LQ models into the six following classes :
($S_{1L}$,$S_3$), ($S_{1R}$,$\tilde{S}_{1}$),  
($R_{2L}$,$R_{2R}$,$\tilde{R}_{2}$), 
($U_{1L}$,$U_{3}$), ($U_{1R}$,$\tilde{U}_{1}$) and 
($V_{2L}$,$V_{2R}$,$\tilde{V}_{2}$).

\subsection{Polarized case}

In a first step, we have tried to pin down the spin asymmetries which 
should allow to disentangle
the chiral structure of the new interaction. Following our previous experience
\cite{PTJMV,JMV} we know that the PV spin asymmetries ($A_L$ or $A_{LL}^{PV}$)
should fulfill this purpose. 
\\

Since their interactions are chiral, the LQs will induce some effects
in these PV asymmetries, and the directions of the deviations 
from the SM expectations
allow the distinction between several classes of models.  
For instance, a positive deviation
for $A_{L}(e^-p)$ pins down the  class ($S_{1L}$,$S_3$) 
(or ($V_{2L}$,$\tilde{V}_{2}$))
and, a negative one, the class ($S_{1R}$,$\tilde{S}_{1}$) (or $V_{2R}$).  
Similarly, an effect for $A_{L}(e^+p)$ makes a
distinction between  the model $R_{2R}$ (or ($U_{1R}$,$\tilde{U}_{1}$))
and the class ($R_{2L}$,$\tilde{R}_{2}$) (or ($U_{1L}$,${U}_{3}$)).  
These properties are illustrated in Fig. 4
which display $A_{L}$ for $e^-p$ and $e^+p$ collisions
with separated plots for scalar and vector LQs. The HERA and LQ parameters
are given in the figure. The statistical and the 10\% systematic
errors are added in quadrature.\\

\begin{figure}
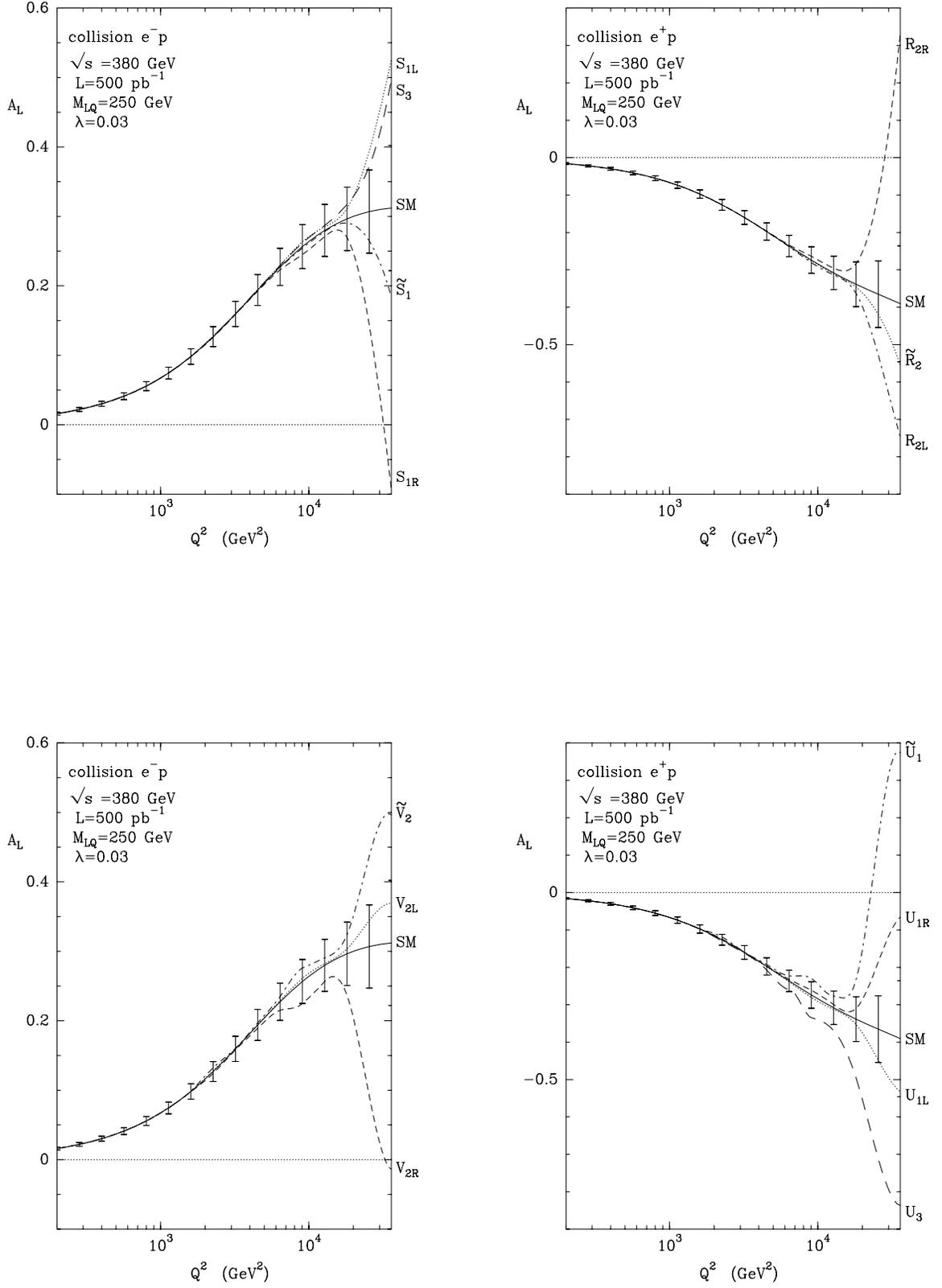

\vspace*{-2.2cm} 
\begin{tabular}[t]{c c}
\centerline{\subfigureA{\psfig{file={fig.4a},width=8truecm,height=12truecm}}
\subfigureA{\psfig{file={fig.4b},width=8truecm,height=12truecm}}}\\
\centerline{\subfigureA{\psfig{file={fig.4c},width=8truecm,height=12truecm}}
\subfigureA{\psfig{file={fig.4d},width=8truecm,height=12truecm}}}
\end{tabular} 
\vspace*{-1.4cm}
\caption{\small $A_L$ vs $Q^2$ for the BRW models.} 
\end{figure}

Therefore, the PV asymmetries separate the 14 BRW models into the following 
eight classes :
($S_{1L}$,$S_3$), ($S_{1R}$,$\tilde{S}_{1}$),  ($R_{2R}$),
($R_{2L}$,$\tilde{R}_{2}$), 
($U_{1L}$,$U_{3}$), ($U_{1R}$,$\tilde{U}_{1}$),
($V_{2L}$,$\tilde{V}_{2}$) and ($V_{2R}$).\\
It appears that the sensitivity of the two-spin PV asymmetry  $A_{LL}^{PV}$
is only slightly better than the one of  $A_{L}$. Therefore, at this step, 
polarized
protons are not mandatory.\\

The final step is to distinguish between an $eu$ and an $ed$ interaction, i.e.
to obtain the flavor of the valence quark involved in the dominant interaction.
With a polarized lepton beam and unpolarized protons,
this flavor separation is not easy since only
the different electric charges or partonic weights of $u$ and $d$ quarks can
be used. Conversely,
when polarized protons are available, it is possible to use a peculiarity of
the polarized valence quark distributions, 
namely $\Delta u > 0$ and $\Delta d <0$ (see e.g. \cite{Reya}
for a recent review).
Indeed, if we pin down a spin asymmetry which is directly proportional
to the $\Delta q$'s, the flavor separation will be obtained from
the sign of the deviation with respect with the SM value for this asymmetry.
The
double spin asymmetries $A_{LL}^{PC}$'s  and the 
polarized charge asymmetries $B_U$ and $B_V$ defined above
share this property. \\
At this point we need to discuss
separately the scalar case and the vector case.\\

\no $\bullet $ {\it Scalar case :} 
\vspace{2.mm}

The three $A_{LL}^{PC}$: $A_1^{PC},A_2^{PC},A_3^{PC}$ are the useful 
observables to separate
the scalar LQs within the remaining classes. 
In Fig. 5 they are displayed
for real LQ production.
\begin{figure}
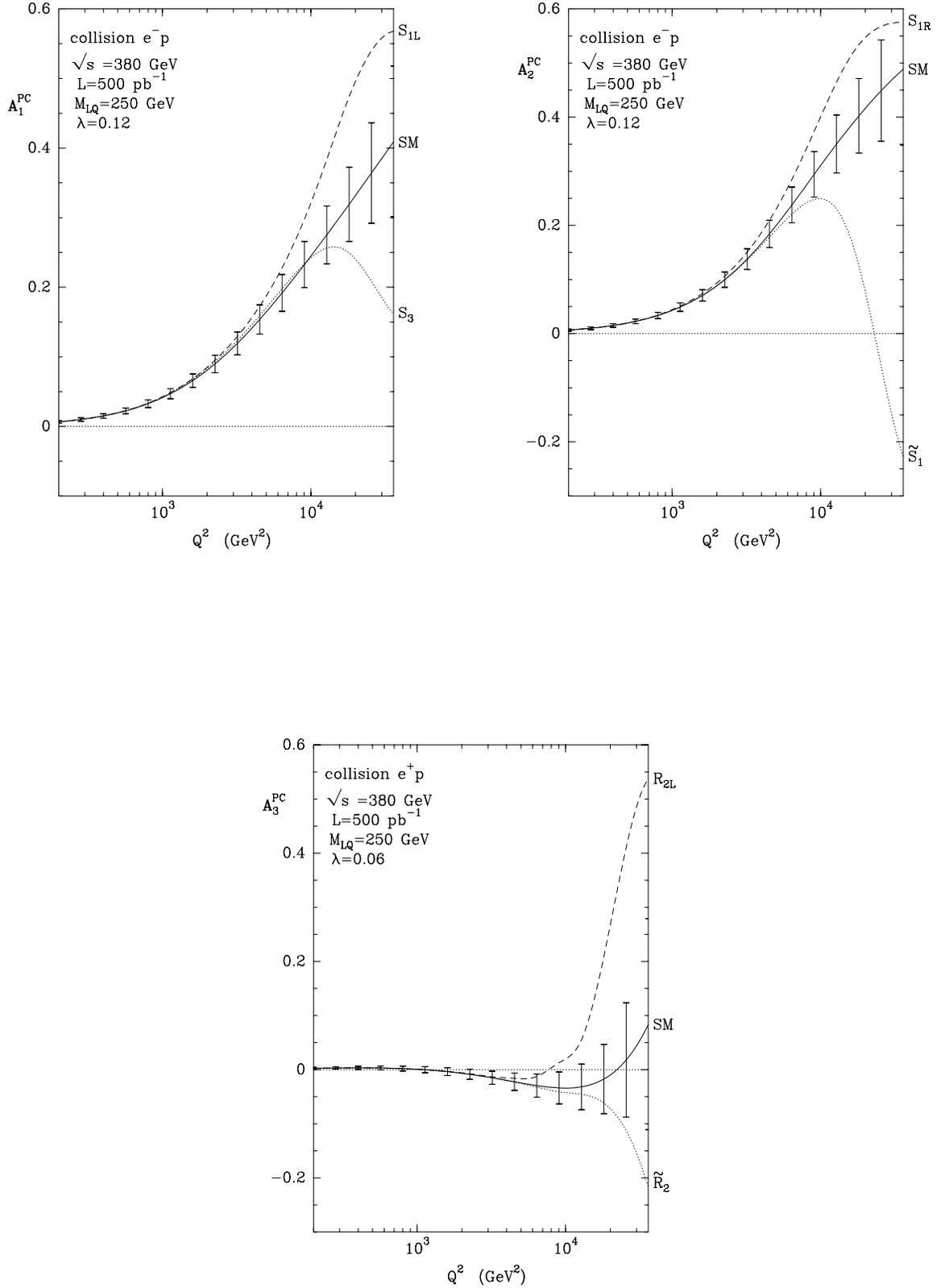

\vspace*{-2.2cm} 
\begin{tabular}[t]{c c}
\centerline{\subfigureA{\psfig{file={fig.5a},width=8truecm,height=12truecm}}
\subfigureA{\psfig{file={fig.5b},width=8truecm,height=12truecm}}}\\
\centerline{\subfigureA{\psfig{file={fig.5c},width=8truecm,height=12truecm}}}
\end{tabular} 
\vspace*{-1.4cm}
\caption{\small $A_{LL}^{PC}$'s vs $Q^2$ for the scalar BRW models.} 
\end{figure}
With the values we have chosen for the LQ parameters the separation is clear.\\

In fact the situation is a little bit more complex. Indeed, at this stage,
we need to know if the LQ is real or virtual in order to pin down the dominant
amplitude.\\
1) If the LQ is on-shell, the dominant term is the squared amplitude for 
LQ production.
This information is known from the observation of the $x$ distribution
of the events \cite{BRW}.\\
2) If the LQ is off-shell, the dominant term is now the ${\gamma }.LQ$ 
interference term.
This information is known from the non-observation of the $s$-resonance.\\
The dominant term controls the behaviour of the asymmetries.
Since it depends on the mass of the LQ, we deduce that the behaviour
of the asymmetries may be also $M$ dependent. In fact we should have a change 
of behaviour
for the LQ models which induce a destructive interference 
with standard $\gamma$ exchange.
For scalar LQs this happens for $\tilde{S}_{1}$, $R_{2R}$ and $R_{2L}$
(see Appendix).\\
At HERA the window for LQ discovery falls essentially in the real domain.
At the TESLAxHERA facility, however,
this distinction between real or virtual LQ exchange will be mandatory.
 
In the following we will consider only the LQs 
in the real domain, at HERA.\\

Therefore, adding  the information which should be obtained from
$A_{L}(e^-p)$ (or from $A_{LL}^{PV}(e^-p)$), we get now a
non-ambiguous separation of the LQ scalar models. This is shown by the different
"deviation signatures" for all the different models presented on table 2.
\begin{center} 
\begin{tabular}{|c||c|c|c|c|c|c|c|} \hline
&$S_{1L}$&$S_{1R}$&$\tilde{S}_{1}$ &$S_3$&$R_{2L}$&$R_{2R}$&$\tilde{R}_{2}$\\  
\hline
\hline $A_{L}(e^-)$& $+$ & $-$ & $-$ & $+$ & $0$ & $0$ & $0$  
\\ \hline
$A_{L}(e^+)$& $0$ & $0$ & $0$ & $0$ & $-$ & $+$ & $-$  
\\ \hline $A_{LL}^{PC} $& $+$ & $+$ & $-$ & $-$ & $+$ & $0$ & $-$ 
\\ \hline 
\end{tabular}
\end{center} 
\begin{center} 
Table 2: ``Deviation signatures'' for the BRW scalar LQ models (real domain).
\end{center} 
\vspace{2.mm}

\no
In this table, "0 deviation" means that the effect of a LQ on a 
particular quantity is contained into the error bar centered on the 
SM expectation. On the other hand, positive and
negative deviations should be clearly visible thanks to the high integrated
luminosities. \\

\no $\bullet $ {\it Vector case :} 
\vspace{2.mm}

Concerning vector LQs, the most sensitive quantities allowing the flavor
separation are the polarized charge asymmetries.
In the real domain the relevant charge asymmetries are $B_U$ and $B_V$.\\
These asymmetries are shown in Fig. 6. We have separated the effects of the classes
($U_{1L}$,$U_{3}$) and ($U_{1R}$,$\tilde{U}_{1}$)
on $B_U$ since these two classes are already
distinguished thanks to the PV asymmetries.
\begin{figure}
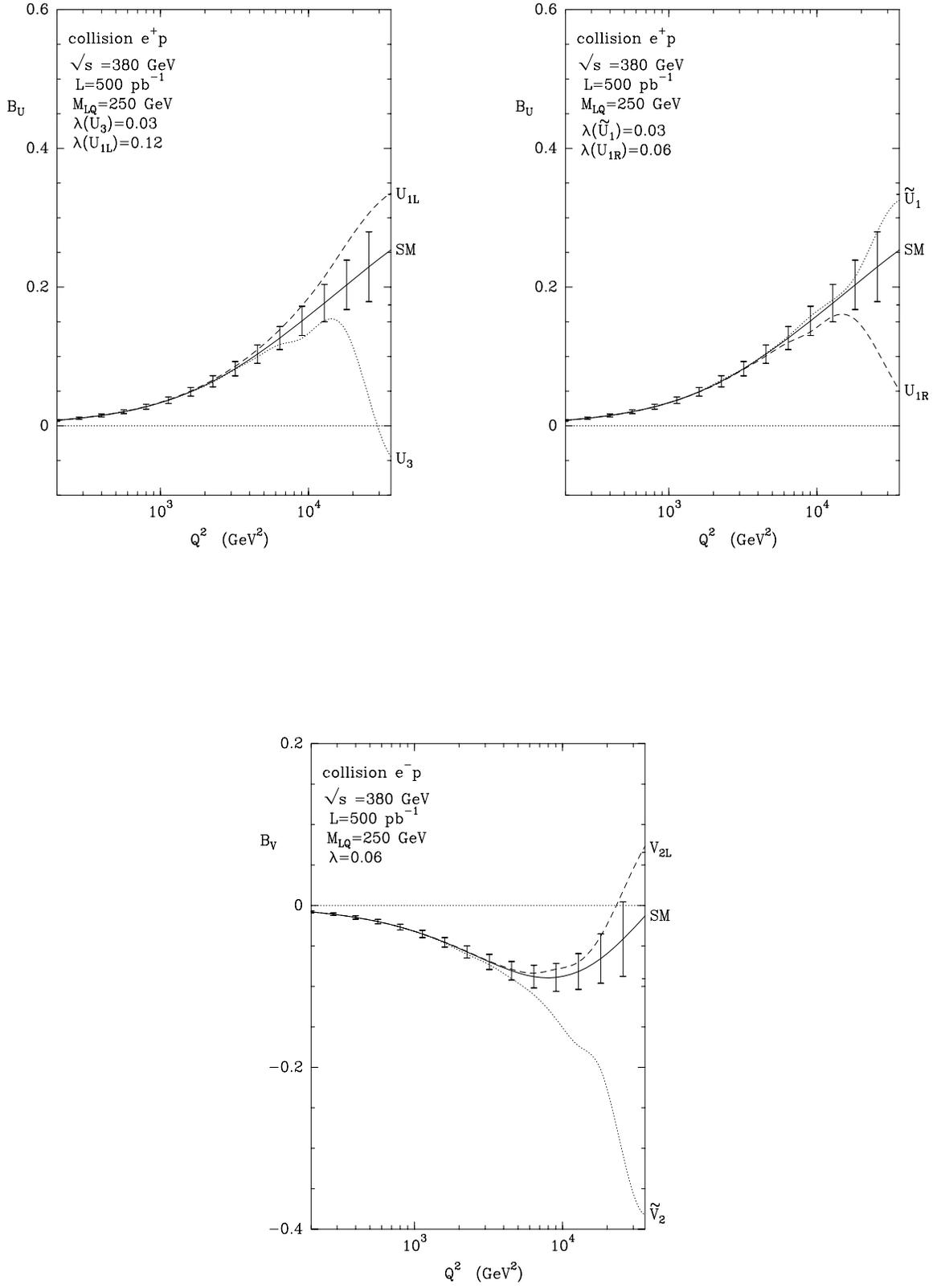

\vspace*{-2.2cm} 
\begin{tabular}[t]{c c}
\centerline{\subfigureA{\psfig{file={fig.6a},width=8truecm,height=12truecm}}
\subfigureA{\psfig{file={fig.6b},width=8truecm,height=12truecm}}}\\
\centerline{\subfigureA{\psfig{file={fig.6c},width=8truecm,height=12truecm}}}
\end{tabular} 
\vspace*{-1.4cm}
\caption{\small $B_U$ and $B_V$ vs $Q^2$ for the vector BRW models.} 
\end{figure}

The deviation signatures for vector LQs are
displayed in table 3.

\begin{center} \begin{tabular}{|c||c|c|c|c|c|c|c|} \hline
&$U_{1L}$&$U_{1R}$&$\tilde{U}_{1}$ &$U_3$&$V_{2L}$&$V_{2R}$&$\tilde{V}_{2}$\\  
\hline
\hline $A_{L}(e^-)$& $-$ & $+$ & $+$ & $-$ & $0$ & $0$ & $0$  
\\ \hline
$A_{L}(e^+)$& $0$ & $0$ & $0$ & $0$ & $+$ & $-$ & $+$  
\\ \hline $B_{U\, or\, V} $& $+$ & $-$ & $+$ & $-$ & $+$ & $0$ & $-$ 
\\ \hline 
\end{tabular}  
\vspace{2.mm}

 Table 3: ``Deviation signatures'' for the BRW vector LQ models (real domain). \\
\end{center}

\subsection{Identification domains}

Finally, it is possible to estimate the domains in the plane ($M, \lambda$)
where a non-ambiguous identification of the nature of a LQ should be
possible. We present in Fig. 7 these ``identification domains'' for some representative
examples.\\
\begin{figure}
\vspace*{-2.2cm} 
\begin{tabular}[t]{c c}
\centerline{\subfigureA{\psfig{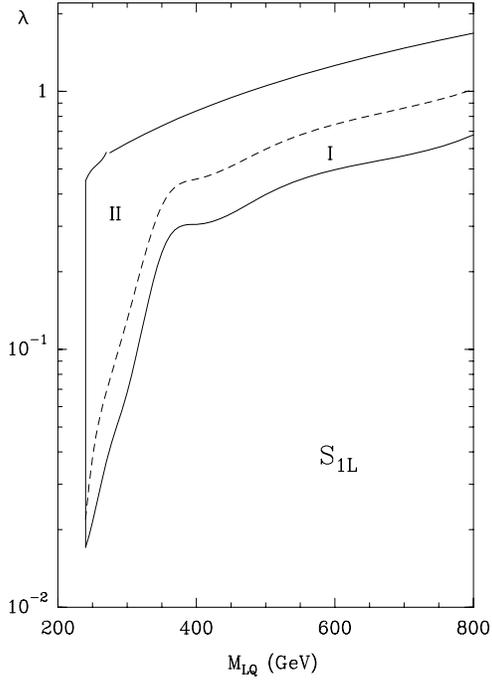}}
\subfigureA{\psfig{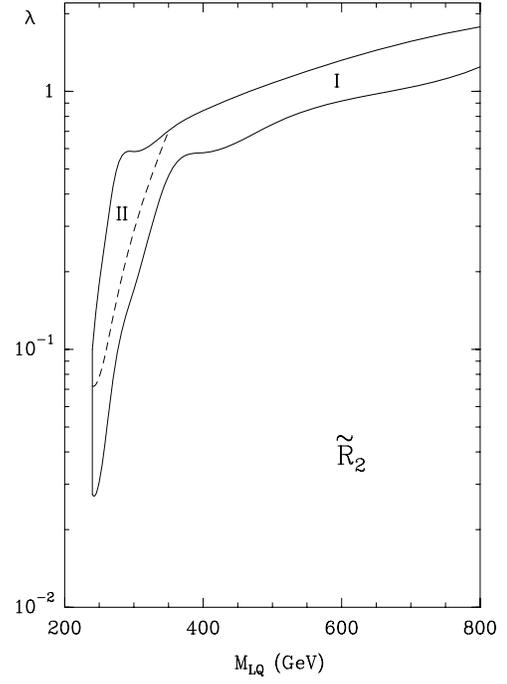}}}\\
\centerline{\subfigureA{\psfig{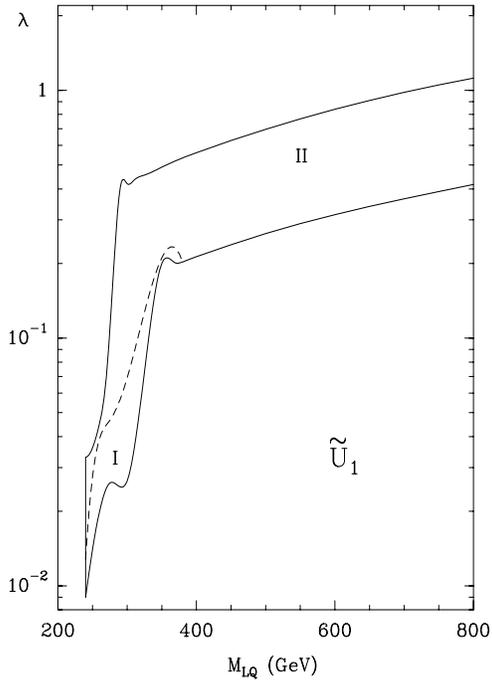}}
\subfigureA{\psfig{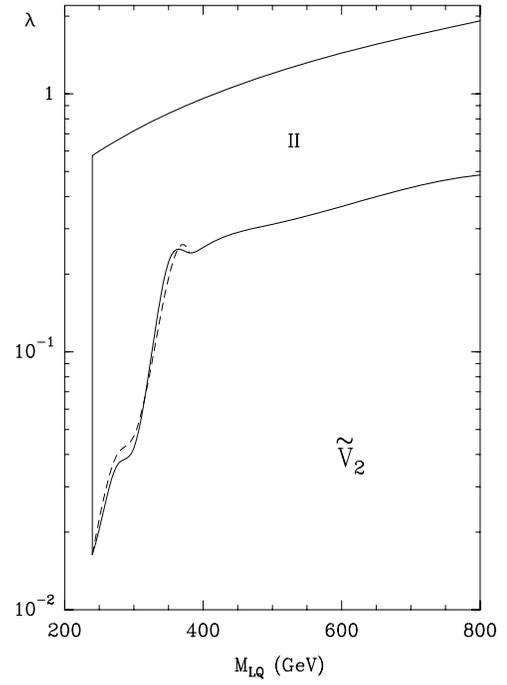}}}
\end{tabular} 
\vspace*{-1.4cm}
\caption{\small Identification domain at 95\% CL for 
${S}_{1L}$, $\tilde{R}_{2}$, $\tilde{U}_{1}$ and $\tilde{V}_{2}$.} 
\end{figure}
The upper curves correspond to the present discovery limits from
Tevatron, HERA and LEP. Constraints from APV have been omitted.
The lower curves represent the constraints
coming from the PV spin asymmetries.
They are better, in general, than the ones from the PC 
or charge asymmetries (dashed curves). 
Note that, for $\tilde{V}_{2}$  
the sensitivities from both types of asymmetries are equivalent.\\
The regions in the parameter 
space where a complete identification of the chiral structure
is possible are given by the domains
I+II. In domain I no effect will appear on the $A^{PC}$'s nor $B$'s and one
misses the flavor separation. In the domain II it is possible to
identify the nature of the LQ without ambiguity.

\medbreak
\section{Conclusion}

Concerning the chances of discovery of Leptoquark states
in the future HERA program (with a high integrated
luminosity), we have seen that there are
still some windows that are not covered by present
data, in particular in the real domain ($M<\sqrt{s}$). 
Measurements of the
integrated unpolarized cross section in NC processes, at the highest
possible energy, should present the best opportunity.
At this stage, polarized beams would not yield better results.\\
Our purpose was mainly to explore the possibilities
of disentangling the various LQ models. 
We present in Fig. 8 a schematic view of what can be done from the 
precise measurements of the various observables we have discussed.\\
The first two steps are well known : with unpolarized $e^-$
and $e^+$ beams it is easy to get in the same time 
the separation between scalars and vectors (from the $y$ distributions)
and between $F=0$ and
$F=2$ LQs (from $d\sigma_{\pm} /dQ^2$).

\m
The next steps are more difficult to perform. However, it is mandatory
to try to pin down the chiral structure of a newly discovered LQ-like
particle. For example it is worth recalling here that, due to SUSY, 
the $R$-parity breaking squarks have universal
left-handed couplings to leptons.

We have shown that polarization of the lepton beam should yield
this information thanks to the precise measurement of $A_L$ in
both $e^-$ and $e^+$ collisions. At this step the polarization
of the proton beams is not necessary.
Note also that the sensitivities
of the PV asymmetry and of the unpolarized cross sections are comparable.
This means that, if polarized
lepton beams are available in the same run, 
as soon as a LQ is discovered in $e^+$ or $e^-$ collisions
(via $d\sigma /dQ^2$), one gets
almost simultaneously his scalar or vector nature (via $d\sigma /dy$)
and the chiral structure of its couplings (via $A_L$) .\\
Now, the next step is to try to get the flavor separation
within the remaining classes of models, which is the most difficult task. 
Indeed, CC processes with
unpolarized beams do not seem to be sufficient to fulfill this program, 
as long as "neutron" beams, through the use
of ionized Deuterium or $^3He$ atoms, are not available.
On the other hand, the behaviours of the polarized 
valence quark distributions $\Delta u$ and $\Delta d$ in a polarized
proton should allow to do this job. In the case
of scalar LQs, measuring the PC double spin asymmetries is sufficient. In the case
of the remaining vectors, it is necessary to measure some polarized
charge asymmetries to obtain the separation at the same level
of sensitivity. 
In both cases, the price to pay is
a proton beam with a high degree of polarization ($P$ = 70\%).
\\

We feel that it was important to get an answer to the following question : 
are both (lepton and proton) polarizations
mandatory to completely disentangle the various LQ models present in
the BRW lagrangians ? According to our analysis the answer is yes.
This conclusion holds certainly also for the TESLA$\times$HERA
project. 

Finally, if we relax the working assumptions {\it i-iv} (see Section 1), 
the LQs can have a more complex
structure and the analysis should be less easy.
In this case, like in the more general
context of Contact Interactions \cite{JMV},
the use of additional asymmetries, that one can also
define with lepton plus proton polarizations, 
should be very useful. \\

Moreover, polarized electron-neutron collisions could be performed
with polarized $^3He$ beams : this option has been seriously
considered in the framework of the RHIC-Spin program at Brookhaven and also
at HERA \cite{hera97}. This could be the final goal of an 
ambitious polarization program at HERA.\\

\begin{figure}[ht]
\hspace*{1.5cm}
    \centerline{\psfig{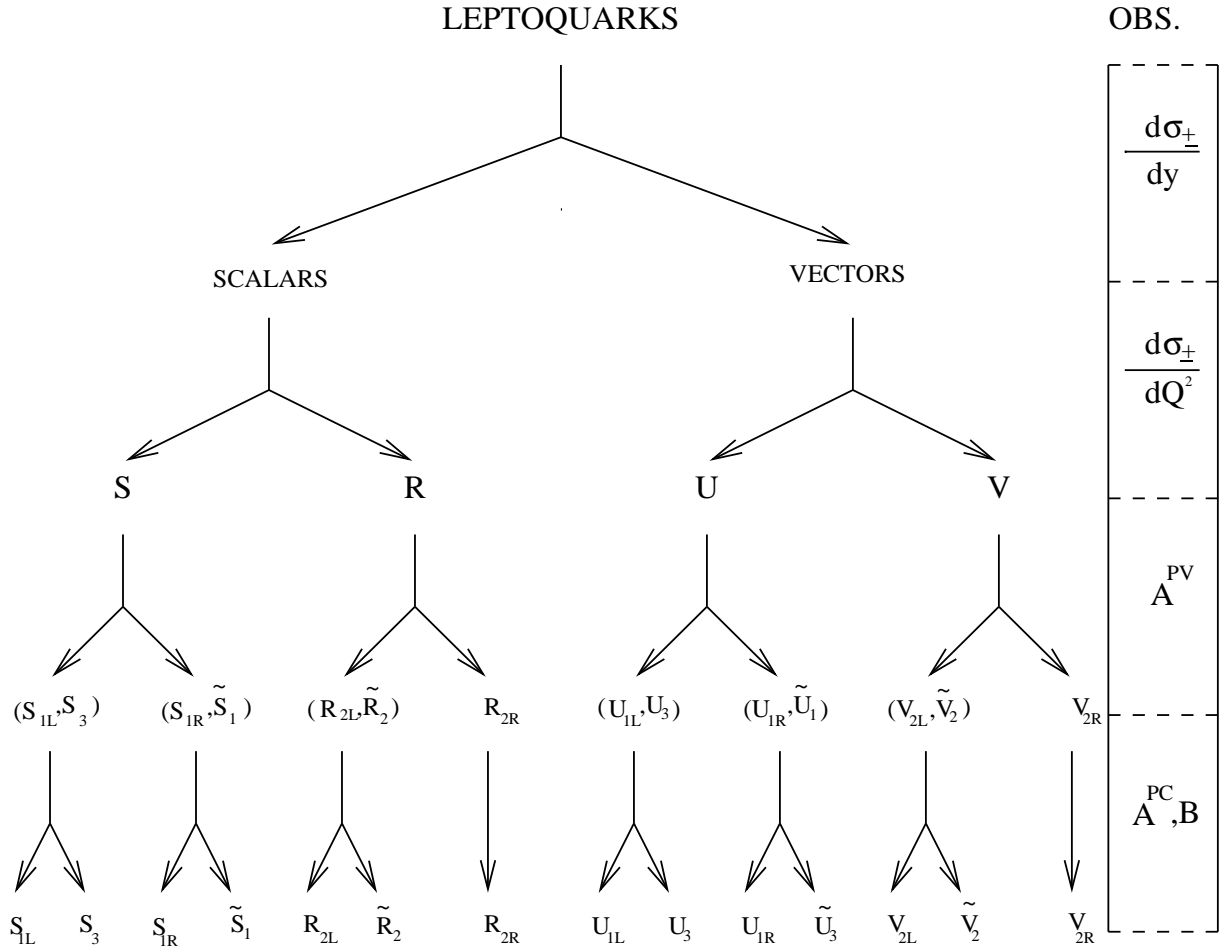}}
    \caption[]{\small Schematic view for LQ identification. }
\end{figure}

\newpage

{\bf Acknowledgements}\\
We are indebted to A. De Roeck, J. Kalinowski, E. Reya, Y. Sirois,
J. Soffer and W. Vogelsang for many interesting discussions and
informations. JMV has the pleasure to thank the theoreticians
of the ``Institut f\"ur Physik'' in Dortmund for helpful
conversations and the excellent atmosphere, and finally, Susanne
Laurent for her kindness and efficiency.
The Alexander von Humboldt foundation is acknowledged for financial support.\\

\vspace{1.truecm}

\newpage

\appendix
\section{Appendix : Cross sections}

\appendixA{}

We present in this appendix, the set of formulas necessary to calculate
the double polarized cross sections, the spin and charge asymmetries
involved in the present analysis.

\subsection{Neutral Current}

\no $\bullet $ {\it \underline{Process}}\\

\no The single polarized cross sections are given in \cite{BRW,kalino}.
Here we give the cross sections in the ($s$, $t$, $u$) notations. \\
\no The collisions between charged leptons and protons, in the neutral
current channel, correspond to the process :
${\vec{e}}^{\,\pm}\; \vec{p}\; \longrightarrow\; e^{\pm}\; X$,
whose cross section is given by :

\EQ\label{defsec}
{\frac{d\sigma_t}{dxdQ^2}}^{h_e h_p}\; =\; 
\sum_{h_q}
{\frac{d\hat{\sigma}_t}{d\hat{t}}}^{h_e h_q}\, q^{h_q}_{h_p}
(x,Q^2)
\eq

\no where $h_e$, $h_p$ and $h_q$ are the helicities
of the charged lepton, proton and parton (quark or antiquark), respectively.
The label $t = \pm$ 
corresponds to the electric charge
of the colliding lepton.
$\sum_q$ represent the sum over all the quark and antiquark flavors
present inside the proton.
The subprocess invariants ${\hat s}$, ${\hat t}$ 
and ${\hat u}$ are given by :
\EQ\label{defsh}
{\hat s} =  x\, s\;\;\;\;\;\;\;\;\;\;\;\;\;\;\;\;\;\;\;\;\;\;\;\; ,
\eq
\EQ
{\hat t} = -Q^2\;\;\;\;\;\;\;\;\;\;\;\;
\;\;\;\;\;\;\;\;\;\; ,
\eq
\EQ\label{defuh}
{\hat u} = x\, u = -x(1-y)\, s \; ,
\eq
\no where the usual variable $y$ is defined by $y=
Q^2/xs$.\\

\no We denote by
$q^{h_q}_{h_p}(x,Q^2)$ the parton distribution for
the parton $q$ inside a proton
of helicity $h_p$, with momentum fraction $x$ and helicity $h_q$, 
at the energy scale $Q^2$. 
These distributions are related to the
parallel and anti-parallel distributions by : $q_+\, =\, q_+^+ \, 
=\, q_-^- \;\; , \;\;
q_-\, =\, q_+^- \, =\, q_-^+ $, which are related to the usual
unpolarized and polarized parton distributions by :
$q\, = \, q_+ + q_-$ and $\Delta q\, = \, q_+ - q_-$.\\


\no $\bullet $ {\it \underline{Subprocesses}}\\

\no Using the notations of \cite{BGS}, the cross section of the elementary
subprocess ${\vec{e}} \,\vec{q} \rightarrow e\, q$ is given by :
\EQ
{\frac{d\hat{\sigma}_t}{d\hat{t}}}^{h_e h_q}\; =
\; \frac{\pi}{{\hat s}^2}\;
\sum_{\alpha , \beta}\; T_{\alpha , \beta}^{h_e h_q}(e^t,q) \; ,
\eq

\no where $T_{\alpha , \beta}^{h_e h_q}(e^t,q)$ is
the squared matrix element for $\alpha$ and $\beta$ boson exchange. $q$
is a quark or an antiquark.
The $T_{\alpha , \beta}^{h_e h_q}(e^t,q)$ for the SM (i.e.
for $\alpha , \beta = \gamma ,\, Z$) are given in \cite{JMV}. 
The $T_{\alpha , \beta}^{h_e h_q}(e^t,q)$ for $LQ$ production, exchange
and interferences with $\gamma$ or $Z$, are given below.
We have
omitted the hat symbol of the variables ${\hat s}$, ${\hat t}$ 
and ${\hat u}$, for clarity.\\

\no {\it\bf Subprocess}\hu {\bf $e^-\; q\; \longrightarrow \; e^-\; q$} :
\begin{eqnarray}
T_{SS}&=&F^2
\frac{1}{64\pi^2}\frac{s^2}{s_S^2+M_S^2\Gamma_S^2}\left[\lambda_L^4
\left(1-h_e\right)(1-h_q)+\lambda_R^4\left(1+h_e\right)\left(1+h_q\right)\right] \\
T_{VV}&=&F^2
\frac{1}{16\pi^2}\frac{u^2}{s_V^2+M_V^2\Gamma_V^2}\left[\lambda_L^4
\left(1-h_e\right)(1+h_q)+\lambda_R^4\left(1+h_e\right)\left(1-h_q\right)\right] \\
T_{RR}&=&F^2
\frac{1}{64\pi^2}\frac{u^2}{u_R^2}\left[\lambda_L^4\left(1-h_e\right)(1
+h_q)+\lambda_R^4\left(1+h_e\right)\left(1-h_q\right)\right] \\
T_{UU}&=&F^2
\frac{1}{16\pi^2}\frac{s^2}{u_U^2}\left[\lambda_L^4\left(1-h_e\right)(1
-h_q)+\lambda_R^4\left(1+h_e\right)\left(1+h_q\right)\right] \\
T_{\gamma S}&=&-F\frac{\alpha Q_eQ_q}{4\pi 
t}\frac{s^2s_S}{s_S^2+M_S^2\Gamma_S^2}\left[\lambda_L^2\left(1-h_e\right)(1-h_q)
+\lambda_R^2\left(1+h_e\right)\left(1+h_q\right)\right] \\
T_{\gamma V}&=&-F\frac{\alpha Q_eQ_q}{2\pi 
t}\frac{u^2s_V}{s_V^2+M_V^2\Gamma_V^2}\left[\lambda_L^2\left(1-h_e\right)(1+h_q)
+\lambda_R^2\left(1+h_e\right)\left(1-h_q\right)\right]
\\
T_{\gamma R}&=&F\frac{\alpha Q_eQ_q}{4\pi 
t}\frac{u^2}{u_R}\left[\lambda_L^2\left(1-h_e\right)(1+h_q)+\lambda_R^2
\left(1+h_e\right)\left(1-h_q\right)\right] \\
T_{\gamma U}&=&F\frac{\alpha Q_eQ_q}{2\pi 
t}\frac{s^2}{u_U}\left[\lambda_L^2\left(1-h_e\right)(1-h_q)+\lambda_R^2
\left(1+h_e\right)\left(1+h_q\right)\right] \\
T_{ZS}&=&-F\frac{\alpha_Z}{4\pi 
t_Z}\frac{s^2s_S}{s_S^2+M_S^2\Gamma_S^2}[\lambda_L^2C_{eL}C_{qL}
\left(1-h_e\right)\left(1-h_q\right)+\lambda_R^2C_{eR}C_{qR}\left(1+h_e\right)
\left(1+h_q\right)] \nonumber\\
\\
T_{ZV}&=&-F\frac{\alpha_Z}{2\pi 
t_Z}\frac{u^2s_V}{s_V^2+M_V^2\Gamma_V^2}[\lambda_L^2C_{eL}C_{qR}
\left(1-h_e\right)\left(1+h_q\right)+\lambda_R^2C_{eR}C_{qL}\left(1+h_e\right)
\left(1-h_q\right)] \nonumber\\
\\
T_{ZR}&=&F\frac{\alpha_Z}{4\pi 
t_Z}\frac{u^2}{u_R}[\lambda_L^2C_{eL}C_{qR}\left(1-h_e\right)\left(1+h_q\right)+
\lambda_R^2C_{eR}C_{qL}\left(1+h_e\right)\left(1-h_q\right)] \\
T_{ZU}&=&F\frac{\alpha_Z}{2\pi 
t_Z}\frac{s^2}{u_U}[\lambda_L^2C_{eL}C_{qL}\left(1-h_e\right)\left(1-h_q\right)+
\lambda_R^2C_{eR}C_{qR}\left(1+h_e\right)\left(1+h_q\right)]  
\end{eqnarray}

\no where $\alpha$ is the electromagnetic coupling,
$\alpha_Z = \alpha/\sin^2 \theta_W \cos^2 \theta_W$, $t_{Z} = 
t-M^2_{Z}$. $C_{fL}$ and $C_{fR}$ are the usual Left-handed and Right-handed
couplings of the $Z$ to the fermion $f$, given by
$C_{fL}=I^f_3 - e_f \sin^2 \theta_W $, $C_{fR}=- e_f \sin^2 \theta_W $
with $I_3^f = \pm 1/2$. For LQs, $s_{LQ} = s-M^2_{LQ}$ and
$u_{LQ} = u-M^2_{LQ}$. 
The values for $\lambda_L^2$, $\lambda_R^2$
and the factor $F$ are given
in table 4 for scalar LQs and in table 5 for vector LQs.
The factor $F$, given in term of combinations of kronecker products, 
is relevant only when we convolute subprocess cross sections
with pdfs.
\begin{center} 
\begin{tabular}{|c||c|c|c|c|c|c|c|} \hline
&$S_{1L}$&$S_{1R}$&$\tilde{S}_{1}$ &$S_3$&$R_{2L}$&$R_{2R}$&$\tilde{R}_{2}$\\  
\hline
\hline $\lambda^2_{L}$& $\lambda^2$ & $0$ & $0$ & $\lambda^2$ & $\lambda^2$ & $0$ & $\lambda^2$  
\\ \hline
$\lambda^2_{R}$& $0$ & $\lambda^2$ & $\lambda^2$ & $0$ & $0$ & $\lambda^2$ & $0$  
\\ \hline $F $& $\delta_{qu}$ & $\delta_{qu}$ & $\delta_{qd}$ 
& $\delta_{qu}+2\delta_{qd}$ & $\delta_{qu}$ & $\delta_{qu}+\delta_{qd}$ & $\delta_{qd}$ 
\\ \hline 
$\lambda_{eq}.\lambda_{\nu q'}$& $-\lambda^2$ & $0$ & $0$ & $+\lambda^2$ & $0$ & $0$ & $0$
\\ \hline 
\end{tabular}
\end{center} 
\begin{center} 
Table 4: Parameters for the BRW scalar LQ models.
\end{center} 
\vspace{2.mm}

\begin{center} \begin{tabular}{|c||c|c|c|c|c|c|c|} \hline
&$U_{1L}$&$U_{1R}$&$\tilde{U}_{1}$ &$U_3$&$V_{2L}$&$V_{2R}$&$\tilde{V}_{2}$\\  
\hline
\hline $\lambda^2_{L}$& $\lambda^2$ & $0$ & $0$ & $\lambda^2$ & $\lambda^2$ & $0$ & $\lambda^2$  
\\ \hline
$\lambda_{R}^2$& $0$ & $\lambda^2$ & $\lambda^2$ & $0$ & $0$ & $\lambda^2$ & $0$  
\\ \hline $F$& $\delta_{qd}$ & $\delta_{qd}$ & $\delta_{qu}$ 
& $2\delta_{qu}+\delta_{qd}$ & $\delta_{qd}$ & $\delta_{qu}+\delta_{qd}$ & $\delta_{qu}$
\\ \hline 
$\lambda_{eq}.\lambda_{\nu q'}$& $+\lambda^2$ & $0$ & $0$ & $-\lambda^2$ & $0$ & $0$ & $0$\\ \hline 
\end{tabular}  
\end{center} 
\begin{center} 
Table 5: Parameters for the BRW vector LQ models. 
\end{center}
\vspace{2.mm}

\no {\it\bf Subprocess}\hu {\bf $e^+\; q\; \longrightarrow \; e^+\; q$} :\\

The squared matrix elements $T_{\alpha ,\beta}$ are obtained from the twelve
preceding equations with the following changes :
$h_e \longleftrightarrow -h_e$, $s \longleftrightarrow u$,
$1/(s_{LQ}^2+M_{LQ}^2\Gamma_{LQ}^2) \longleftrightarrow 1/u_{LQ}^2$
and $s_{LQ}/(s_{LQ}^2+M_{LQ}^2\Gamma_{LQ}^2) \longleftrightarrow 1/u_{LQ}$.\\

\no {\it\bf Subprocess}\hu {\bf $e^{\pm}\; {\bar q}\; \longrightarrow \; 
e^{\pm}\; {\bar q}$} :\\

The $T_{\alpha ,\beta}$ are obtained from the ones for $e^{\mp} q$
scattering after the  same transformations as above plus
$h_q \longleftrightarrow -h_q$.
\\

\subsection{Charged Current}

The process for CC is : ${\vec{e}}^{\,\pm}\; \vec{p}\; 
\longrightarrow\; \nu (\bar{\nu })\; X$\\
All the preceding formulas hold for CC processes with the 
following substitutions :
$Q_{e,q} \rightarrow  0$,
$C_L \rightarrow 1$,
$C_R \rightarrow 0$,
$\alpha_Z \rightarrow \alpha_W=\alpha_Z\cos^2 \theta_W$,
$M_Z \rightarrow M_W$ and $t_Z \rightarrow t_W$.
Concerning factor $F$, we have $F=\delta_{qu}\; (\delta_{qd})$ for
$S_{1L}$ and $S_3$ ($U_{1L}$ and $U_3$).
Finally, for $W.LQ$ interferences one has now two different vertices
in the diagram for $LQ$ exchange (i.e. $e.q.LQ$ + ${\nu}.q'.LQ$ vertices).
Then the squared coupling $\lambda^2 \equiv \lambda^2_{eq}$ appearing
in NC, is changed to the product $\lambda_{eq}\lambda_{\nu q'}$.
From eqs.1,2 or from table 1 of \cite{kalino} we have
$\lambda_{eq}=\pm\lambda_{\nu q'}=\pm\lambda$. 
The product $\lambda_{eq}\lambda_{\nu q'}$ is given in the last row of tables 4 and 5.

\end{document}